\documentclass[11pt]{article}
\usepackage{graphicx}
\usepackage{amsmath}
\usepackage{dsfont}
\usepackage{amsfonts}
\usepackage{filecontents}
\usepackage{slashed}
\usepackage{microtype}
\usepackage[macrosonly]{chet}
\usepackage{xcolor}
\definecolor{darkblue}{rgb}{0.1,0.1,.7}
\definecolor{darkgreen}{rgb}{0,.5,0}
\usepackage[pdftex,breaklinks,colorlinks,linkcolor=darkblue,
citecolor=darkgreen]{hyperref}

\newcommand{\cO}{{\cal O}}

\renewcommand{\b}{\beta}

\newcommand{\m}{\mu}
\newcommand{\n}{\nu}

\newcommand{\bphi}{\bar{\phi}}

\def\be{\begin{equation}}
\def\ee{\end{equation}}
\def\bea{\begin{eqnarray}}
\def\eea{\end{eqnarray}}
\def\ba{\begin{array}}
\def\ea{\end{array}}
\def\bi{\begin{itemize}}
\def\ei{\end{itemize}}

\newcommand{\beq}{\begin{equation}}
\newcommand{\eeq}{\end{equation}}
\newcommand{\beqn}{\begin{eqnarray}}
\newcommand{\eeqn}{\end{eqnarray}}
\newcommand{\bga}{\begin{align}}

\def\dalemb#1#2{{\vbox{\hrule height .#2pt
\hbox{\vrule width.#2pt height#1pt \kern#1pt
\vrule width.#2pt}
\hrule height.#2pt}}}

\def\b{\beta}

\def\bz{{\overline{z}}}

 \def\IZ{\relax\ifmmode\mathchoice
 {\hbox{\cmss Z\kern-.4em Z}}{\hbox{\cmss Z\kern-.4em Z}}
 {\lower.9pt\hbox{\cmsss Z\kern-.4em Z}}
 {\lower1.2pt\hbox{\cmsss Z\kern-.4em Z}}\else{\cmss Z\kern-.4em Z}\fi}
 \def\IB{\relax{\rm I\kern-.18em B}}
 \def\IC{{\relax\hbox{$\inbar\kern-.3em{\rm C}$}}}
 \def\Ic{{\relax\hbox{$\inbar\kern-.22em{\rm c}$}}}
 \def\ID{\relax{\rm I\kern-.18em D}}
 \def\IE{\relax{\rm I\kern-.18em E}}
 \def\IF{\relax{\rm I\kern-.18em F}}
 \def\IG{\relax\hbox{$\inbar\kern-.3em{\rm G}$}}
 \def\IGa{\relax\hbox{${\rm I}\kern-.18em\Gamma$}}
 \def\IH{\relax{\rm I\kern-.18em H}}
 \def\II{\relax{\rm I\kern-.18em I}}
 \def\IK{\relax{\rm I\kern-.18em K}}
 \def\IP{\relax{\rm I\kern-.18em P}}

 \font\cmss=cmss10 \font\cmsss=cmss10 at 7pt
 \def\IR{\relax{\rm I\kern-.18em R}}

\def\cO{{\cal O}}

\def\b{\beta}

\def\m{\mu}
\def\n{\nu}

\def\cC{{\cal C}}
\def\cQ{{\cal Q}}

\newcommand{\thistitle}{ \huge \bf{Thermal one-point functions and single-valued polylogarithms} }

\begin{document}

\allowdisplaybreaks
\title{\thistitle}
\author{\Large Anastasios C. Petkou
	\\
	\\
	{\small Division of Theoretical Physics,
	School of Physics }\\
	{\small Aristotle University of Thessaloniki} \\
	{\small 54124 Thessaloniki, Greece.}
	
	\\
	}
\date{\today}
\maketitle
\vspace{-5ex}


\begin{abstract}

I point out that the thermal one-point functions of a pair of relevant operators in massive free QFTs, in odd dimensions and in the presence of an imaginary chemical potential for a $U(1)$ global charge, are given by certain classes of single-valued polylogarithms. This result is verified by a direct calculation using the thermal OPE. The complex argument of the polylogarithms parametrize a two-dimensional subspace of relevant deformations of generalised free CFTs, while the rank of the polylogarithms is related to the dimension $d$. This may be compared with the well-known representation of single-valued polylogarithms as multiloop Feynman amplitudes. As an example, the thermal one-point function of the $U(1)$ charge in $d$-dimensions generalises the thermal average of the twist operator in a pair of harmonic oscillators and is given by the well-known conformal ladder graphs in four dimensions. 
\end{abstract}

\maketitle

\section{Introduction.}
The coupling of critical systems to background geometries is at the heart of many interesting questions in QFT and string theory. Loosely speaking, a nontrivial gravitational background probes operator deformations in the space of CFTs and that leads to a deep relationship between renormalised background effective actions and QFT correlation functions giving many interesting results in the recent past 
\cite{Osborn:1991gm,Osborn:1993cr,Komargodski:2011vj}. 

A particularly simple yet nontrivial background is the thermal geometry
$S^1_{\beta}\times\mathbb{R}^{d-1}$
that is used to describe field theories at finite temperature $T=1/\b$, with $\b$ the inverse circumference of the circle. For critical systems described by CFTs, the main effect of the thermal background is to induce nontrivial one-point functions for quasiprimary operators $\cO(x)$ with scaling dimension $\Delta_{\cO}$ that behave schematically as $\langle \cO(x)\rangle_{\beta}\propto b_\cO/\beta^{\Delta_{\cO}}$. On the other hand thermal one-point functions are intimately related to  the thermal free energy of the theory since the latter is essentially the effective action on the thermal background.  In a remarkable application of this idea one finds that the central charge of two-dimensional CFTs determines the leading correction to temperature scaling of the thermal free energy of the corresponding critical system \cite{Cardy:1986ie,Bloete:1986qm}. In $d>2$ there is no notion of CFT central charge but the relationship between the temperature scaling of the free energy and the thermal one-point function of the energy-momentum tensor still holds true \cite{Cardy:1987dg,Petkou:1998fb,Petkou:1998fc}. 

Perhaps the simplest examples of critical systems are free CFTs in general $d$-dimensions. Their RG deformations at finite temperature have been extensively studied using standard thermal field theory techniques (e.g. see \cite{Laine:2016hma} for a review) and  more recently using the thermal conformal OPE \cite{Iliesiu:2018fao,Petkou:2018ynm}. In this short note, combining old results and new calculations I point out that the thermal free energy of massive free QFTs in the presence of an imaginary chemical potential for a global $U(1)$ charge is given by a particular class of single-valued polylogarithms constructed by F. Brown  \cite{BROWN2004527} (For reviews see \cite{Schnetz:2013hqa,Todorov:2014tya}). The free QFTs considered here may be viewed as deformations  of generalised free CFTs by a pair of relevant operators corresponding to a mass term and a $U(1)$ charge. The latter case is equivalent to considering a charged system in the grand canonical ensemble with imaginary chemical potential.\footnote{See \cite{Kashiwa:2019ihm} for a recent review on imaginary chemical potential techniques in field theory.} I show that the one-point function of the $|\phi|^2$ operator corresponding to the massive deformation in $d$-dimensions is proportional to the thermal free energy in $d-2$-dimensions. I also show that the  one-point functions of the deforming operators in $d$ and $d+2$ dimensions are related through differential equations which are exactly the ones satisfied by single-valued polylogarithms  As a particular example, I point out that the thermal one-point functions of the $U(1)$ charge operator is given in terms of the celebrated conformal ladder graphs in $d=4$ \cite{Usyukina:1992jd,Usyukina:1993ch}. Among the possible implications of our observations one may conceive a deep connection between multidimensional thermal partition functions and conformal gauge theories in $d=4$.

\section{Thermal one-point functions from the grand canonical ensemble in free field theories}
In the standard imaginary time formalism for thermal fields theories\footnote{The Euclidean variables are defined as $x^\m=(\tau,\bar{x})$. $\tau\in [0,\b]$ and the bosonic fields obey periodic boundary conditions $\phi(t+\b,\vec{x})=\phi(\tau,\vec{x})$.} the Euclidean action  action for a massive complex scalar field $\phi(x)$ in odd $d$-dimensions in the presence of an imaginary chemical potential (or equivalently in the presence of the temporal component of real gauge potential of a Chern-Simons gauge field see e.g. \cite{Filothodoros:2016txa,Filothodoros:2018pdj}) is \cite{Laine:2016hma}
\be
\label{SE}
{\cal S}_E(\b;m,\m)=\int_0^\beta d\tau\int d^{d-1}\vec{x} \,\,|(\partial_\tau-i\m)\phi|^2+|\vec{\partial}\phi|^2+m^2|\phi|^2\,.
\ee
To evaluate the grand canonical partition function and the corresponding free energy (grand canonical potential) of the theory one may cure the short distance singularity and other calibration issues by subtracting the zero temperature mass and chemical potential results to obtain
 \be
\label{Zgcphi}
Z_{gc}(\b;m,\m)\equiv\frac{1}{Z(0;0,0)}\int ({\cal D}\bphi)({\cal D}\phi)e^{-S_E}=e^{-\b F_{gc}(\b;m,\m)}\,.
\ee
By a simple scaling argument the grand canonical free energy is usually written as
\be
\label{Fgc}
F_{gc}(\b;m,\m)=\frac{V_{d-1}}{\b^{d}}{\cC}_d(\b m,\b\m)\,,
\ee
with $V_{d-1}$ the spatial volume. The dimensionless function ${\cC}_d$ has been extensively studied over the years e.g. \cite{CastroNeto:1992ie,Appelquist:1999hr,LeClair:2005aa} as a measure of the degrees of freedom along the RG, although it does not appear to satisfy the requirements of a $c$-theorem \cite{Klebanov:2011gs}. 

The properties of $\cC_d$ as a function of the parameters $m$, $\m$ and notably also $d$, are nicely presented using the set of complex variables 
\be
\label{zzbar}
z=e^{-\b m-i\b\m}\,,\,\,\,\bz =e^{-\b m+i\b\m}\,\Rightarrow\,\cC_d(\b m,\b\m)\equiv\cC_d(z,\bz)\,.
\ee
The calculation of $F_{gc}$ can be done along the lines described in the appendix of \cite{Filothodoros:2018pdj} and yields
\be
\label{IntFgc}
\frac{1}{V_{d-1}}F_{gc}=\frac{1}{2}\int\frac{d^{d}p}{(2\pi)^{d}}\ln\frac{p^2+m^2}{p^2}+\frac{1}{\b}\int\frac{d^{d-1}\vec{p}}{(2\pi)^{d-1}}\Re\ln\left(1-e^{-\b\sqrt{\vec{p}^2+m^2}-i\b\m}\right)\,.
\ee
The final result can be brought into the form
\be
\label{fd}
\cC_d(z,\bz)=-K_d\ln^d|z|-\frac{S_{d-1}}{(2\pi)^{d-1}}\left[i_d(z,\bz)+\bar{i}_d(z,\bz)\right]\,,\,
\ee
where 
\be
\label{KdSd}
K_d=\frac{\pi S_d}{d(2\pi)^d}\frac{1}{\sin(\pi d/2)}\,,\,\,S_{d}=\frac{2\pi^{d/2}}{\Gamma(d/2)}\,.
\ee
For $d\geq 3$ the integral $i_d(z,\bz)$ is given by
\be
\label{id}
i_d(z,\bz)=\int_0^z\frac{dw}{w}\left(\ln w-\frac{1}{2}\ln\frac{z}{\bz}\right)\left[\left(\ln w-\frac{1}{2}\ln\frac{z}{\bz}\right)^2-\ln^2|z|\right]^{\frac{d-3}{2}}\!\!\!\!\!\!\!\ln(1-w)\,,
\ee
and $\bar{i}(z,\bz)$ is obtained from (\ref{id}) by exchanging $z\leftrightarrow \bz$. For $d=1$ we have $i_1(z)=-\ln(1-z)$ and $\bar{i}_1(\bz)=-\ln(1-\bz)$. For odd $d$ the calculation of $\cC_d$ can be reduced to a finite series of iterated integrals and then to a finite double series using results such as
\be
\label{intcalc}
\int_0^z\frac{dw}{w}\left(\ln w-\frac{1}{2}\ln\frac{z}{\bz}\right)^k \ln(1-w)=\sum_{\ell=0}^{k}\frac{(-1)^{\ell+1}k!}{\ell !}\ln^\ell|z|Li_{k+2-\ell}(z)\,,
\ee
where $Li_n(z)$ are the usual polylogarithms. At the end one obtains
\be
\label{intresult}
i_d(z,\bz)+\bar{i}_d(z,\bz)=-\frac{\Gamma\left(\frac{d+1}{2}\right)}{d-1}I_{d+2}(z,\bz)\,,
\ee
with
\be
\label{Id}
I_d(z,\bz)=\sum_{n=0}^{\frac{d-3}{2}}\frac{(-1)^n(d-3-n)!}{\left(\frac{d-3}{2}-n\right)!}\frac{2^n\ln^n|z|}{n!}\left[Li_{d-2-n}(z)+Li_{d-2-n}(\bz)\right]\,.
\ee
In the simple free field theory studied here $m^2$ and $\m$ parametrize the deformations of the free Hamiltonian by the operators  $\cO=|\phi|^2$ and $\cQ=i\bphi\overleftrightarrow{\partial_\tau}\phi$, the latter being the charge density corresponding to the $U(1)$ current $J_\m = i\bphi\overleftrightarrow{\partial_\m}\phi$ \cite{Laine:2016hma}. Evidently, the normalized thermal averages (integrated thermal one-point functions) of the above deformations are obtained as moments of the free energy. Assuming they are uniform\footnote{Namely, $\int \langle O(x)\rangle =\b V_{d-1}\langle O\rangle$.} one obtains
\begin{align}
\label{Oaverage}
\langle\cO\rangle_d&=\frac{1}{\b^{d-2}}\hat{\bf D}\cC_d(z,\bz)\,,\,\,\,\hat{\bf D}=\frac{1}{2\ln|z|}(z\partial_z+\bz\partial_\bz)\,,\\
\label{Qaverage}
\langle\cQ\rangle_d&=\frac{1}{\b^{d-1}}\hat{\bf L}\cC_d(z,\bz)\,,\,\,\,\hat{\bf L}=(z\partial_z-\bz\partial_\bz)\,.
\end{align}
We thus see that the thermal one-point functions $\langle\cO\rangle_d$ and $\langle\cQ\rangle_d$ are the responses of $\cC_d$  along the radial and angular directions\footnote{For example, setting $z=\rho e^{i\phi}$ one finds that $2\hat{\bf D}=\partial_\rho$ and $\hat{\bf L}=-i\partial_\phi$.} in the two-dimensional space of massive and $U(1)$ deformations of generalised free CFTs. 

To further unveil the physical content of the relations (\ref{Oaverage}) and (\ref{Qaverage}) it is instructive to consider the case $d=1$. This corresponds to a system of two noninteracting harmonic oscillators with frequency $\omega\equiv m$. The twisted partition function of the oscillators is given by
\be
\label{Z1}
{\cal Z}_1=Tr_{{\cal H}_{1,2}}e^{-\b\hat{H}+i\b\m\hat{\cQ}}\,.
\ee
The Hamiltonian and the twist operator are respectively\footnote{I set $\hbar=1$ and define the creation/anihilation operators as usual 
$\hat{a}^\dagger_i=\frac{1}{\sqrt{2}}(m\hat{x}_i+i\hat{p}_i)\,,\,\,\,\hat{a}_i=\frac{1}{\sqrt{2}}(m\hat{x}_i-i\hat{p}_i)\,,\,\,\,[\hat{x}_i,\hat{p}_j]=i\delta_{ij}\,,\,\,\,i=1,2$.}
\be
\hat{H}=\sum_{i=1}^2\frac{\hat{p}_i^2}{2}+\frac{m^2 x_i^2}{2}\,,\,\,\hat{\cQ}_i=\hat{a}_i^\dagger\hat{a}_i\,\text{(no summation)} \,,\,\,\,\hat{\cQ}=\hat{\cQ}_1-\hat{\cQ}_2\,,
\ee

The trace in (\ref{Z1}) is taken over the tensor product Hilbert space ${\cal H}_{1,2}\approx \{|n_1\rangle\otimes|n_2\rangle\}$, $n_1,n_2=0,1,2,..$. Calculating the partition function one easily obtains
\be
\label{Cd1}
\cC_1(z,\bz)=-\ln|z|+\ln(1-z)+\ln(1-\bz)\,,
\ee
where I used the definitions (\ref{zzbar}). This matches the result (\ref{fd}) for $d=1$. Next, writing the partition the partition function in a slightly unconventional form as
\be
\label{Z1unc}
{\cal Z}_1=Tr_{{\cal H}_{1,2}}e^{-\b(\hat{H}_0+m^2\hat{\cO})+i\b\m\hat{\cQ}}\,,\,\,\,\hat{\cO}=\frac{1}{2}(\hat{x}^2_i+\hat{x}^2_2)\,,
\ee
where $\hat{H}_0=(\hat{p}^2_1+\hat{p}^2_2)/2$ is the {\it free} Hamiltonian or equivalently in this case the kinetic energy, 
one realises that the moments of $\cC_1$ give the thermal averages of the operators $\hat{O}$ and $\hat{\cQ}$ as
\begin{align}
\label{Oaverage1}
\langle\hat{\cO}\rangle_1 &=-\frac{\b}{2\ln|z|}\left[1+|z|^2\left(\frac{1}{z(1-\bz)}+\frac{1}{\bz(1-z)}\right)\right]\,,\\
\label{Qaverage1}
\langle\hat{\cQ}\rangle_1 &=|z|^2\left(\frac{1}{z(1-\bz)}-\frac{1}{\bz(1-z)}\right)\,.
\end{align}
Also recall that in our simple model there is a virial theorem at work relating the thermal averages of the operator $\hat{O}$ and the Hamiltonan $\hat{H}$ as
\be
\label{virial}
2\omega^2\langle\hat{\cO}\rangle_1=\langle\hat{H}\rangle\,.
\ee

The generic formulae (\ref{Oaverage}) and (\ref{Qaverage}) yield an intimate connection among theories in different dimensions. The crucial point is the following result which can be proven by a direct calculation
\be
\label{cdcd2}
\hat{\bf D}\cC_d(z,\bz)=-\frac{1}{4\pi}\cC_{d-2}(z,\bz)\,,
\ee
for $d=1,3,5,..$ with the boundary condition $\cC_{-1}(z,\bz)=-\frac{4\pi}{\b}\langle\cO\rangle_1$. Then, from (\ref{Oaverage}) and (\ref{cdcd2}) one obtains
\be
\label{CdOd2}
\cC_d(z,\bz)=-4\pi\b^{d}\langle\cO\rangle_{d+2}\,,
\ee
which shows that the free energy of the $d$-dimensional theory is proportional to the thermal one-point function of the operator $\cO$ in $d+2$ dimensions. One also finds
\be
\label{OdOd2}
\langle\cO\rangle_d=-4\pi\b^2\hat{\bf D}\langle\cO\rangle_{d+2}\,,
\ee
which connects the thermal one-point functions of the operator $\cO=|\phi|^2$ in free theories across generic odd dimensions. Moreover, since the differential operators $\hat{\bf L}$ and $\hat{\bf D}$ commute one finds
\be
\label{QdQd2}
\langle\cQ\rangle_d=-4\pi\b^2\hat{\bf D}\langle\cQ\rangle_{d+2}\,,
\ee
for $d=1,3,5,..$.

Although our results take a nice form in terms of the complex variables $z$ and $\bz$, it is also interesting to consider the formula (\ref{cdcd2}) just in terms of the variable $m$. It is not hard to rewrite it formally as an iterated integral 
\be
\label{iterated}
\cC_{2n+1}(m_n,\m)=(-4\pi)^n\int^{m_n}\!\!\!\!m_{m-1}dm_{n-1}..\int^{m_1}\!\!\!\!m_0dm_0\,\,\cC_1(m_0,\m)\,,
\ee
where we set $d=2n+1$. The chemical potential $\m$ does not enter in this calculation. Thus we see that the higher dimensional thermal free energies are generated by the free energy of the $d=1$ harmonic oscillator.

\section{Thermal one-point functions from the OPE}
Part of the motivation of this investigation comes from recent work on thermal one-point functions of CFTs in odd dimensions \cite{Iliesiu:2018fao,Petkou:2018ynm}.  
For a complex scalar $\phi(x)$ with dimension $\Delta_\phi=d/2-1$ and periodic boundary conditions the Euclidean $x$-space\footnote{For
the corresponding momentum-space expression see \cite{Petkou:1998fb,
Petkou:1998fc}.} two-point function takes the generic form (we set $\b=1$ for simplicity in this section. The temperature scaling can be trivially restored at the end of the calculations.)
\be
\label{2ptf}
\langle\bphi(x)\phi(0)\rangle\equiv g(r,\cos\theta)=
\sum_{{\cO}_s}a_{\cO_s}r^{\Delta_{\cO_s}}
\frac{C_s^{\nu}(\cos\theta)}{r^{d-2}}\,,
\ee
where 
$r=|x|$ and $\theta\in[0,\pi]$ is a polar angle on $\mathbb{R}^{d-1}$. $C_{s}^{\nu}(\cos\theta)$ are Gegenbauer
polynomials with $\nu=d/2-1$ and (\ref{2ptf}) runs over all
operators ${\cO}_s$ in the OPE $\bphi\times\phi$ with spin $s$ and dimension
$\Delta_{\cO_s}$. The coefficients $a_{\cO_s}$ are related to the thermal one-point functions of the operators in the OPE. In our conventions
\be
\label{aOsbOs}
a_{\cO_s}=\frac{s!}{2^s(\nu)_s}
\frac{g_{\phi\phi\cO_s}b_{\cO_s}}{C_{\cO_s}}\,,\,\,\,\,\langle {\cO}_s(x)\rangle=b_{\cO_s}\left(e_{\m_1}...e_{\m_s}-{\rm traces}\right)\,,
\ee
where $C_{\cO_s}$ and $g_{\phi\phi\cO_s}$ are the corresponding two- and
three-point function coefficients, $e_\m=(1,\vec{0})$ is the unit vector in the $\tau$-direction and $(a)_n$ the Pochhammer symbols. 

In \cite{Petkou:2018ynm} we have studied the thermal two-point function of the elementary scalars in a theory corresponding to mass deformed free CFT in general odd dimensions $d=2k+1$, $k=1,2,..$. Using the Euclidean inversion formula of \cite{Iliesiu:2018fao} we determined the coefficients $a_{{\cO}_s}$ by evaluating the residues at the physical poles of a suitable spectral function. We found a spectrum that consists of three classes of operators: i) higher-spin operators with dimensions $\Delta_s=d-2+s$, $s=0,2,4,..$ along with their higher-twist generalizations, ii) scalar operators with dimensions $\Delta_l=d-2+2l$, $l=0,1,2..$ and iii) scalar {\it shadow} operators with dimensions $\Delta_k=2k$, $k=1,2,..$.  

In the case of a real scalar field studied in \cite{Petkou:2018ynm} $\phi^2$ with dimension $\Delta_{\phi^2}=d-2$ was the only relevant operator in the OPE $\phi\times\phi$ which was {\it not a  shadow operator}, and there were no currents with odd spin. In the case of the complex scalar field studied here there is another relevant operator which is not a shadow operator; this is the $s=1$ current $J_\m$ with dimension $\Delta_J=d-1$. From (\ref{2ptf}) one can write the contribution of the above two operators in the thermal two-point function as
\be
\label{2ptfphiJ}
g(r,\cos\theta)=\frac{a^{}_{\mathbb{I}}}{r^{d-2}}+[{\rm shadows}]+\frac{g^{}_{\phi\phi\phi^2}}{C^{}_{\phi^2}}b^{}_{\phi^2}+r\cos\theta\frac{g^{}_{\phi\phi J}}{C^{}_J}b^{}_J+\cdots
\ee
The relevant OPE coefficients depend on the thermal mass and the imaginary chemical potential, namely $b_{{\cO}_s}=b_{{\cO}_s}(z,\bz)$. In view of the discussion above we see that the operators $\phi^2$ and $J_\tau$ correspond to the operators $\cO$ and ${\cal Q}$ whose thermal one-point functions are given by the general formulae (\ref{Oaverage}) and (\ref{Qaverage}).  If one requires that (\ref{2ptfphiJ}) is a conformal OPE, then the arbitrary scale parameter introduced in the theory by $b_{\phi^2}$ should not be there. Requiring its vanishing is the condition (gap equation) suggested in \cite{Petkou:2018ynm} that determines the critical values of the thermal mass and chemical potential. For $\m=0$ this condition gives a class of  algebraic equations for high order polylogarithms and it explains the critical behaviour of bosonic systems in generic even dimensions.\footnote{See \cite{Giombi:2019upv} for a related result in $d=5$.} It would be nice to study these equations further for $\m\neq 0$ and to see whether there is a second condition involving $b_J$. 

 The thermal two-point function of the free theory (\ref{SE}) is \cite{Iliesiu:2018fao,Petkou:2018ynm}
 \be
 \label{thermal2ptf}
 g(r,\cos\theta;m,\m)=\frac{1}{(2\pi)^{\frac{d}{2}}}\sum_{n=-\infty}^\infty e^{in\m}\left(\frac{m}{|X_{n,\zeta}|}\right)^{\n}K_\n\left(m|X_{n,\zeta}|\right)\,,
\ee
with $X_{n,\zeta}=\sqrt{(n-\zeta)(n-\bar{\zeta})}$, $\zeta=re^{i\theta}$, and $K_\n(y)$ is a modified Bessel function with half integer index. To bring (\ref{thermal2ptf}) into the form (\ref{2ptfphiJ}) we note that the $n=0$ term gives the usual zero temperature result for the massive two-point function, while the Bessel functions have polynomial expansions for odd $d$. In $d >2$ the zero temperature contribution has a divergent  expansion as $m\rightarrow 0$ where two cut-off independent terms stand out: the term giving the massless two-point function and the $r$-independent term proportional to $m^{d-2}$. In the expansion (\ref{2ptfphiJ}) these correspond to the contribution of the unit and $|\phi|^2$ operators respectively. Notice that the contributions from shadow scalar operators come from the process of the cut-off subtraction and renormalization of the theory which is implicity assumed here and it does not affect the results. After some algebra one obtains
\begin{align}
\label{bphi2}
\frac{g_{\phi\phi\phi^2}}{C_{\phi^2}}b_{\phi^2}&=a_{\mathbb{I}}\frac{\Gamma\left(\frac{d-1}{2}\right)}{\Gamma(d-2)}\left[\frac{\Gamma\left(1-\frac{d}{2}\right)}{2\sqrt{\pi}}m^{d-2}+I_d(z,\bar{z})\right]\,,\\
\label{bJ}
\frac{g_{\phi\phi J}}{C_J}b_J&=-\frac{1}{2}a_{\mathbb{I}}\frac{\Gamma\left(\frac{d-1}{2}\right)}{\Gamma(d-2)}{\cal I}_d(z,\bar{z})\,,
\end{align}
where $
a_{\mathbb{I}}=\sqrt{\frac{\pi}{2}}\frac{1}{(2\pi)^{\frac{d}{2}}}\frac{1}{2^{\frac{d-3}{2}}}\frac{\Gamma(d-2)}{\Gamma\left(\frac{d-1}{2}\right)}$. The  function $I_d(z,\bar{z})$ was given in (\ref{Id}) and ${\cal I}_d(z,\bz)$ is
\be
\label{calI}
{\cal I}_d(z,\bar{z}) =\sum_{n=0}^{\frac{d-1}{2}}\frac{(-1)^n(d-1-n)!}{\left(\frac{d-1}{2}-n\right)!}\frac{2^n\ln^n|z|}{n!}\left[Li_{d-1-n}(z)-Li_{d-1-n}(\bz)\right]\,.
\ee
Taking into account the explicit results for the two- and three-point function coefficients\footnote{Notice that with our definitions $g_{\phi\phi J}$ is imaginary.} one can verify that (\ref{bphi2}) and (\ref{bJ})  are just another form of the formulae given in the previous section.

\section{Thermal one-point functions as single-valued polylogarithms.}

Our results (\ref{fd}), (\ref{intresult}) and \ref{Id}) are particular cases of the single-valued polylogarithms $P_w(z,\bz)$ constructed by Brown \cite{BROWN2004527}. One way to see this is to rewrite (\ref{cdcd2}) is the form
\be
\label{LaplaceEq}
\partial_z\partial_\bz \cC(z,\bz)=-\frac{d-1}{8\pi}\frac{1}{|z|^2}\cC_{d-2}(z,\bz)\,,
\ee
which is valid for $d\geq 3$, while for $d=1$ we use the boundary condition given below (\ref{cdcd2}). Using then (\ref{Qaverage}) and (\ref{Qaverage1}) and the index $n=(d-1)/2$, $n=1,3,..$ we find that the thermal one-point functions of the $U(1)$ charge satisfy the sequence of differential equations (we set $\b=1$ here)
\be
\label{sequence}
\partial_z\partial_\bz \langle\cQ\rangle_{n=2}=\frac{1}{4\pi}\left(\frac{1}{\bz(1-z)}-\frac{1}{z(1-\bz)}\right)\,,\,\,\,\partial_z\partial_\bz \langle\cQ\rangle_{n+1}=-\frac{n}{4\pi}\frac{1}{|z|^2}\langle\cQ\rangle_n\,,\,\,\text{for}\,\,n=2,3,..
\ee
These are exactly the differential equations studied in \cite{BROWN2004527} and shown to have as unique single-valued solutions the celebrated result of the ladder graphs \cite{Usyukina:1992jd,Usyukina:1993ch}. 

Explicitly one observes that for $d\geq 1$ (see eq. 2.38 of \cite{Schnetz:2013hqa})
\be
\label{idSVP}
I_d(z,\bz)=(-1)^{\frac{d+1}{2}}\Gamma\left(\frac{d+1}{2}\right)P_{w_d}(z,\bz)\,,\,\,\,w_d\equiv 0^{\{n_d\}}10^{\{n_d\}}\,,
\ee
with $n_d=(d-3)/2$. The index $w_d$ denotes a "word" in the "two-letter alphabet" $\{0,1\}$ and one normalises such that $P_{\emptyset}=1$ and $P_{0^{\{n\}}}=2^n\ln^n|z|/n!$.  Notice that $P_1(z,\bz)=\ln(1-z)+\ln(1-\bz)$. It can be shown that the single-valued polylogarithm $P_{w_d}(z,\bz)$ satisfies for $d\geq 1$
\be
\label{KZeqs}
z\partial_zP_{0w_d0}(z,\bz)=P_{0w_d}(z,\bz)\,,\,\,\,\bz\partial_\bz P_{0w_d0}(z,\bz)=P_{w_d0}(z,\bz)\,,
\ee
which are of the form of Knizhnik-Zamolodhikov equations studied in \cite{BROWN2004527}. Actually, one can show that the second order differential equations implied by relations such as (\ref{cdcd2}) give rise to the differential equations studied e.g. in \cite{Drummond:2012bg}.
The KZ-like equations (\ref{KZeqs}) then imply
\begin{align}
\label{LPwd}
\hat{\bf L}P_{0w_d0}(z,\bz)&=P_{0w_d}(z,\bz)-P_{w_d0}(z,\bz)\,\\
\label{DPwd}
\hat{\bf D}P_{0w_d0}(z,\bz)&=\frac{1}{n_d+1}P_{w_d}(z,\bz)=\frac{2}{d+1}P_{0w_{d-2}0}(z,\bz)\,,
\end{align}
which are effectively a different form of the relationships (\ref{cdcd2}), (\ref{OdOd2}) and (\ref{QdQd2}).

\section{Outlook}
It is well-known that single-valued polylogarithms are closely related to Feynman graphs in gauge and string theories. The observation made here seems to connect two apparently  different physical quantities; thermal one-point functions in odd-dimensional theories and multiloop graphs in ${\cal N}=4$ SYM in $d=4$. Contrasting these two manifestations of single-valued polylogarithms one notices that in the first case the variables $z$ and $\bz$ parametrise relevant deformations of free CFTs while in the second case parametrize spacetime points. Moreover, in the first case the order of the polylogarithms is related to the spacetime dimension $d$, while in the latter case to the number of loops in the graphs.

 It would be nice to understand further the implications, if any, of our observation. As an example one may consider interpreting the charge one-point functions (\ref{Qaverage}) as higher-dimensional generalizations of the thermal average of the twist operator in the system of two harmonic oscillators. For example, using (\ref{fd}) and e.g. (\ref{LPwd}) one finds for $d=3$ (see also \cite{Filothodoros:2018pdj})
\be
\label{Q3}
4\pi\b^2\langle\cQ\rangle_3=-4iD(d)\,,\,\,D(z)=\Im Li_2(z)+\ln|z| {\rm Arg}(1-z)\,,
\ee 
with $D(z)$ the famous Bloch-Wigner function. It is well-known that $D(z)$ appears in the simplest tree graph in a four-dimensional CFT
\be
 \label{I111}
 I(x_1,x_2,x_3,x_4)=\frac{1}{\pi^2}\int\frac{d^4 x}{(x-x_1)^2(x-x_2)^2(x-x_3)^2(x-x_4)^2}=\frac{1}{x_{14}^2x_{23}^2}\frac{4i}{z-\bar{z}}D(z)\,,
 \ee
 where the change of variables is now
 \be
 \label{vuzz0}
  u=\frac{x_{12}^2x_{34}^2}{x_{13}^2x_{24}^2}\,,v=\frac{x_{12}^2x_{34}^2}{x_{14}^2x_{23}^2}=(1-z)(1-\bar{z})\,,\,\,\,\frac{v}{u}=z\bar{z}\,.
 \ee
What is perhaps less known is that the integral (\ref{I111}) can be obtained as a limit that correponds to a particular kind of twisting. Namely, using the result given in eq. B.12 of \cite{Petkou:1994ad} (and in many other places since then e.g. \cite{Hogervorst:2016itc,Giombi:2018vtc}), one obtains after some algebra 
\be
\label{GblockPhi1}
\lim_{\epsilon\to 0}\Gamma(\epsilon)\left[G^{(4)}_{2-\epsilon}(v,u)-C(4,2-\epsilon)v^\epsilon G_{2+\epsilon}^{(4)}(v,u)\right]=\frac{4i}{z-\bar{z}}D(z)\,,
\ee
where $G_\Delta^{(d)}(v,u)$ is the standard $d$-dimensional conformal block of a scalar operator with dimension $\Delta$ and
\be
\label{C}
C(d,\Delta)=\frac{\Gamma(\Delta)\Gamma^4(d/2-\Delta/2)\Gamma(1-d/2+\Delta)}{\Gamma(d-\Delta)\Gamma^4(\Delta/2)\Gamma(1+d/2-\Delta)}\,.
\ee
The important point, first stressed in \cite{Hoffmann:2000tr,Hoffmann:2000mx}, is that the $\epsilon\rightarrow 0$  divergence cancels out in (\ref{GblockPhi1}) leaving a finite result.\footnote{The cancellation process in $d=4$ was explicitly given by Dolan \& Osborn in \cite{Dolan:2000uw,Dolan:2000ut}.} Hence, the first nontrivial ladder graph gives a twisting of shadow scalar conformal blocks. Perhaps this interpretation generalises to higher loops. 

One may also suspect the relevance of our results to recent works on large-charge expansions where imaginary chemical potentials play an important role \cite{Alvarez-Gaume:2019biu}. In another direction, it would be interesting to study the manifestation of the shuffle algebra of the polylogarithms in the context of thermal field theories.

\section*{Acknowledgements}
I would like to thank A. Stergiou and C. Wen for many useful discussions on this and related ongoing work. I also thank L. Ciambelli, R. G. Leigh,  M. Petropoulos and O. Schnetz for their interest and relevant discussions.
This work was supported by the Hellenic Foundation for Research and Innovation (H.F.R.I.) under the ``First Call for H.F.R.I. Research Projects to support Faculty members and Researchers and the procurement of high-cost research equipment grant" (MIS 1524, Project Number: 96048).

\bibliographystyle{ieeetr}
\bibliography{PolylogRefs}

\end{document}